\documentclass[pdflatex,sn-nature,iicol]{sn-jnl}
\usepackage{graphicx}%
\usepackage{multirow}%
\usepackage{amsmath,amssymb,amsfonts}%
\usepackage{amsthm}%
\usepackage{mathrsfs}%
\usepackage[title]{appendix}%
\usepackage{xcolor}%
\usepackage{textcomp}%
\usepackage{manyfoot}%
\usepackage{booktabs}%
\usepackage{algorithm}%
\usepackage{algorithmicx}%
\usepackage{algpseudocode}%
\usepackage{listings}%
\usepackage{lineno}
\usepackage{cuted}
\usepackage{caption}

\theoremstyle{thmstyleone}%
\theoremstyle{thmstyletwo}%
\theoremstyle{thmstylethree}%
\renewcommand{\thesubsection}{\arabic{subsection}}
\begin{document}
\title[Article Title]{Compact plasma-based extreme compression of relativistic electron beams}

\author[1]{\fnm{R.} \sur{Hessami}}
\equalcont{These authors contributed equally to this work.}
\author[1]{\fnm{K.K.} \sur{Swanson}}
\equalcont{These authors contributed equally to this work.}
\author[2]{\fnm{M.} \sur{Galletti}}
\equalcont{These authors contributed equally to this work.}
\author[2]{\fnm{L.} \sur{Verra}}
\equalcont{These authors contributed equally to this work.}
\author[2]{\fnm{A.} \sur{Biagioni}}
\author[2]{\fnm{L.} \sur{Crincoli}}
\author[2,3]{\fnm{M.} \sur{Carpanese}}
\author[2]{\fnm{M.} \sur{Del Giorno}}
\author[2]{\fnm{F.} \sur{Demurtas}}
\author[2]{\fnm{G.} \sur{Griseri}}
\author[2]{\fnm{V.} \sur{Lollo}}
\author[2,3]{\fnm{M.} \sur{Opromolla}}
\author[2,3]{\fnm{F.} \sur{Nguyen}}
\author[2]{\fnm{G.} \sur{Parise}}
\author[2,3]{\fnm{A.} \sur{Petralia}}
\author[2]{\fnm{G.J.} \sur{Silvi}}
\author[2]{\fnm{F.} \sur{Stocchi}}
\author[2]{\fnm{M.} \sur{Zottola}}
\author*[1]{\fnm{A.} \sur{Marinelli}}
\author*[2]{\fnm{R.} \sur{Pompili}}
\equalcont{These authors contributed equally to this work.}
\author*[1]{\fnm{C.} \sur{Emma}}\email{cemma@slac.stanford.edu, riccardo.pompili@lnf.infn.it, marinelli@slac.stanford.edu}
\equalcont{These authors contributed equally to this work.}

\affil[1]{\orgname{SLAC National Accelerator Laboratory}, \orgaddress{\street{2575 Sand Hill Road}, \city{Menlo Park}, \postcode{94025}, \state{California }, \country{USA}}}
\affil[2]{\orgname{INFN - Laboratori Nazionali di Frascati}, \orgaddress{\street{Via Enrico Fermi 54}, \postcode{00044}, \city{Frascati},  \country{Italy}}}
\affil[3]{ENEA Nuclear Department, R.C. Frascati, Via Enrico Fermi 45, 00044 Frascati, Italy}

\abstract{Accelerated particle beams are essential tools of modern science and technology with applications ranging from medicine~\cite{Durante2017}, to light sources~\cite{lightsources_org}, to high-energy physics~\cite{ShiltsevZimmermann2021ModernFutureColliders}.  Over time, as accelerator-based experiments become more sophisticated and probe natural phenomena with greater temporal resolution, there is an increasing demand on beam parameters, in particular on bunch duration. Experiments have been proposed that require beams with 1--2 orders of magnitude shorter durations than what is possible with current state-of-the-art technology. However, state-of-the-art bunch compressors force a trade-off between how short an electron bunch can be made and how well its transverse quality survives the strong bending magnets required to do so — a limitation that stands as a grand challenge for accelerator physics to-date~\cite{Blazey2023ABPRoadmap}. In this experiment, we demonstrate a new, compact compression technique for generating ultrashort, high-peak-current electron bunches that circumvents this trade-off using plasma wakefield acceleration. Just as chirped pulse amplification unlocked ultrashort, high-power laser pulses and the technologies that followed from them, here we demonstrate an analogous chirp-and-compress strategy applied to relativistic electron beams by using the extreme electric fields of a plasma wakefield in place of a stretcher-amplifier, thereby unlocking the path to ultrashort, extremely high-current electron beams. We demonstrate the first application of this plasma-compressed bunch, using it to drive single-microbunch free electron lasing at optical wavelengths in a single, meter-scale undulator module, producing radiation pulses roughly four orders of magnitude more energetic than the first beam-driven, plasma-based free electron laser demonstration at the same accelerator facility~\cite{Pompili2022,Galletti2024}.}

\maketitle

\clearpage
\begin{strip}
 \centering
\includegraphics[width=\textwidth]{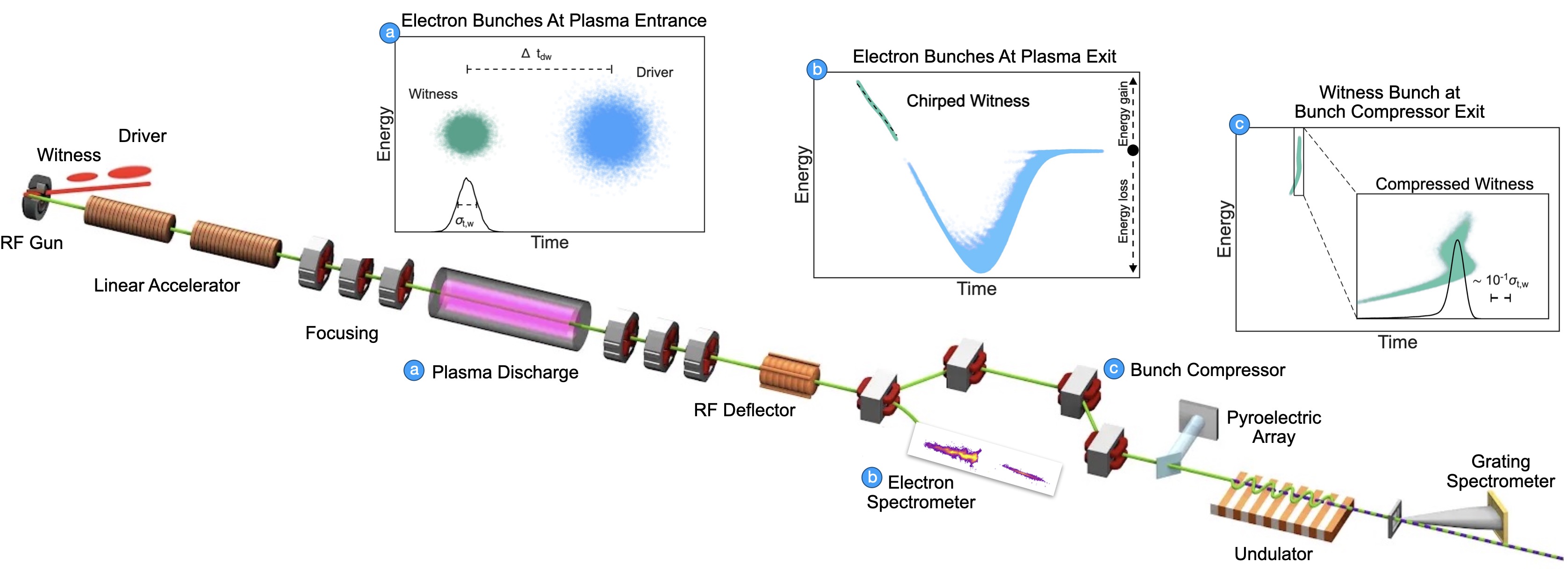}
\captionof{figure}{\textbf{Experimental layout.} Schematic longitudinal phase space of the driver and witness bunches at (a) the plasma entrance, (b) at the plasma exit, and (c) downstream of the compression chicane. The compressed witness bunch radiates coherently in a downstream undulator tuned to a resonant wavelength of 650~nm comparable to the compressed witness bunch length.
}\label{Figure_1}
\end{strip}

\par Relativistic electron beams are one of the cornerstones of modern science, serving as invaluable tools for studying nature at its most fundamental level~\cite{SesslerWilson2014}. Continued progress in our ability to improve the quality of electron beams (that is, focus them more tightly in space and compress them to shorter durations in time) has led to an increasingly detailed understanding of matter down to the angstrom spatial and sub-femtosecond temporal scales \cite{Hemsing2014}. Despite this continued progress, next-generation experiments across disparate fields such as strong-field quantum electrodynamics~\cite{Matheron2023}, ultrafast quantum and materials science~\cite{Cesar2022,Kim2026}, laboratory astrophysics~\cite{Benedetti2018} and attosecond science \cite{berrah2025attosecond} will require an order of magnitude leap in electron bunch compression compared to the current state of the art. However, it has long been known that traditional bunch compression techniques force a trade-off: extreme temporal compressions are usually achieved at the expense of a large deterioration of the transverse beam quality due, for instance, to coherent synchrotron radiation (CSR) emission~\cite{Talman1986, Nakazato1989,Saldin1997,Braun2000,DiMitri2013}. Overcoming this limitation in the face of ever-increasing demands on beam properties for experiments has been recognized as one of the grand challenges facing accelerator physics today~\cite{Blazey2023ABPRoadmap}.

In this work, we present the first experimental demonstration of a novel and compact temporal compression technique of an electron bunch using a scalable method that alleviates such a trade-off by exploiting the strong accelerating fields in a plasma-wakefield accelerator~\cite{Swanson2026}.
Similarly to how the chirped pulse amplification technique~\cite{Strickland1985,Strickland2019} paved the way for the generation of ultrashort high-power laser pulses, with ensuing revolutionary applications across society~\cite{Maine1988, Mourou2006, Mourou2019}, our methodology enables the generation of orders of magnitude shorter electron bunches, potentially reaching the attosecond scale with mega-ampere peak currents. This, in turn, translates to equivalent peak electric fields in the multi-teravolt-per-meter range. These electron bunches will be the tools advancing the frontiers of knowledge of the future, driving instruments such as carrier envelope phase stable, few-cycle, waveform-controlled extreme ultraviolet (EUV)/X-ray free electron lasers (FELs)~\cite{Tibai2014,Emma2021,Hessami2024,Morgan2026} and enabling breakthrough experiments in ultrafast quantum and materials science~\cite{Cesar2022,Kim2026}, strong-field quantum electrodynamics~\cite{Matheron2023} and beyond.

What enables extreme compression of the electron beam in this experiment is the extremely large electric fields that can be sustained in a plasma accelerator (orders of magnitude larger than with typical radiofrequency (RF) cavity-based accelerators)~\cite{Esarey2009,Lindstrom2025}. These ultra-strong electric fields allow one to imprint strong longitudinal energy chirps on the beam. The large-amplitude chirp is key to the extreme compression mechanism for two reasons: first, the shortest achievable bunch duration for a given beam brightness is inversely proportional to the energy chirp; second a large chirp allows one to compress the electron beam without using strong bends~\cite{Emma2021}, which typically spoil the electron beam quality in traditional compressor designs. In our experiment, the chirped bunch is then compressed in a weakly bending, downstream magnetic chicane and its bunch duration is reduced by an order of magnitude. This fully compressed electron bunch is then utilized in the first demonstration of single-microbunch free electron lasing at optical frequencies in a compact, meter-scale, downstream undulator~\cite{Gover2019}. In the following, we describe in detail the beam-plasma interaction which generates the witness chirp, the temporal compression mechanism and the bunch lasing in an FEL. The experimental results are further validated and compared with results from numerical simulations. 

\begin{figure}[!t]
    \centering
    \includegraphics[width=1\linewidth]{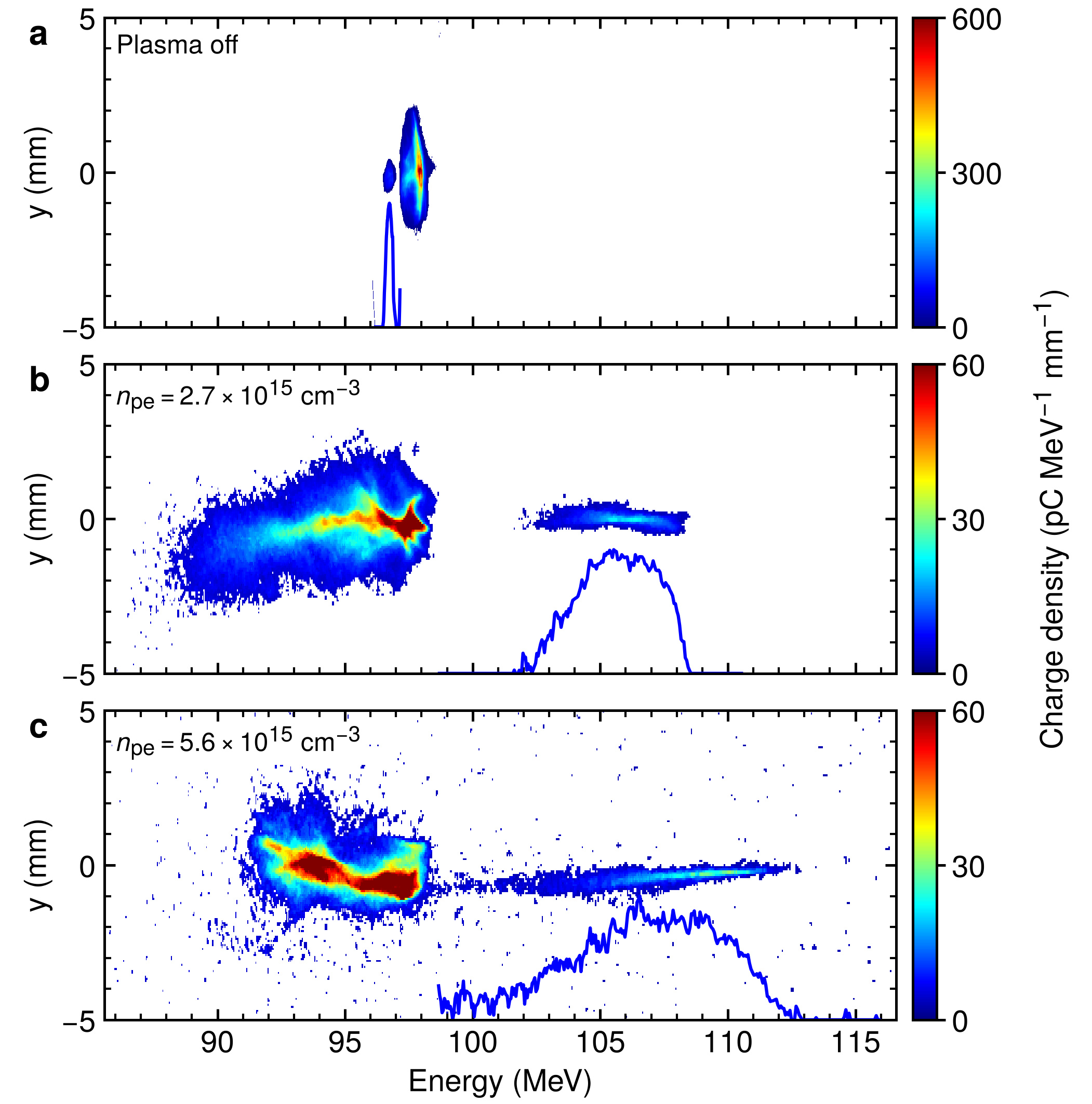}
    \caption{\textbf{Electron energy spectra.} Single-event energy spectra of the driver and witness bunches (a) without plasma, (b) with plasma at density $2.7\times10^{15}$~cm$^{-3}$, and (c) at density $5.6\times10^{15}$~cm$^{-3}$. The blue lines show the energy distribution of the witness bunch.}
    \label{Figure_2}
\end{figure}

The experimental setup is shown in Fig.~\ref{Figure_1}. Two electron bunches, hereinafter called \textit{driver} and \textit{witness}, are generated by the SPARC\_LAB photoinjector, which consists of an RF-gun followed by three accelerating sections~\cite{Pompili2026}. The bunches are generated by the photoemission of a copper cathode illuminated by two blue laser pulses~\cite{pompili2021time} and are accelerated and longitudinally compressed using the velocity-bunching technique~\cite{serafini2001velocity,Ferrario2011}. The resulting longitudinal phase space (LPS) at the plasma entrance consists of a driver and witness bunch of energy $\sim$96~MeV, a separation on the order of 1~ps and bunch charges of 500~pC and 20~pC respectively (Fig.~\ref{Figure_2}a). The driver and witness bunch durations are estimated via RF deflector measurements and beam dynamics simulations to be 500~fs and 30~fs RMS, respectively.
The two electron bunches are then focused to an RMS transverse spot size of $\sim$50~$\mu$m into a 13-cm-long, high-voltage, discharge plasma capillary by means of a triplet of permanent magnet quadrupoles (PMQs)~\cite{pompili2018compact}. 
The entire setup is installed in a vacuum chamber directly connected with a windowless, three-stage differential pumping system that ensures $10^{-8}$~mbar pressure in the RF linac while flowing gas into the capillary. This solution allows beam transport through the capillary without encountering any window, thus not degrading the beam emittance by multiple scattering~\cite{pompili2018focusing}.
The plasma density in the discharge capillary is adjusted from $10^{15}$ to $10^{16}$~cm$^{-3}$ by changing the delay between the beginning of the discharge and the arrival of the electron bunches. More details are reported in Methods.

As the drive bunch propagates through the plasma, it excites a large-amplitude wakefield. The witness bunch is placed in the accelerating phase of the wakefield where the amplitude of the longitudinal wakefield increases linearly along the bunch. This results in the witness bunch being accelerated and simultaneously acquiring a large energy chirp as it transits through the plasma. As shown in Fig.~\ref{Figure_2}, the large energy chirp manifests as an increase in RMS energy spread of the witness bunch in the electron energy spectrum. 
In the absence of beam loading and when the plasma wakefield interaction is in the nonlinear regime~\cite{Rosenzweig1991}, the energy chirp acquired by the witness beam is proportional to the plasma density~\cite{Swanson2026}. We demonstrate the tunability of the energy chirp by measuring the LPS of the witness bunch at two different plasma densities, with representative shots shown in Fig.~\ref{Figure_2}b,c. At a plasma density of $2.7\times10^{15}$~cm$^{-3}$, the witness is accelerated to a mean energy of 106.1~MeV and acquires a $1.6 \pm 0.2$~MeV RMS projected energy spread. At this density, the range of RMS projected energy spreads measured over 100 consecutive shots is 1.03--2.86~MeV. At a plasma density of $5.6\times10^{15}$~cm$^{-3}$, the witness is accelerated further up to 109.9~MeV, acquiring a $4.5 \pm 0.5$~MeV RMS energy spread. The range of RMS projected energy spreads measured over 100 consecutive shots is 1.8--5.01~MeV. 
 
\begin{figure}[!t]
\centering
\includegraphics[width=\linewidth]{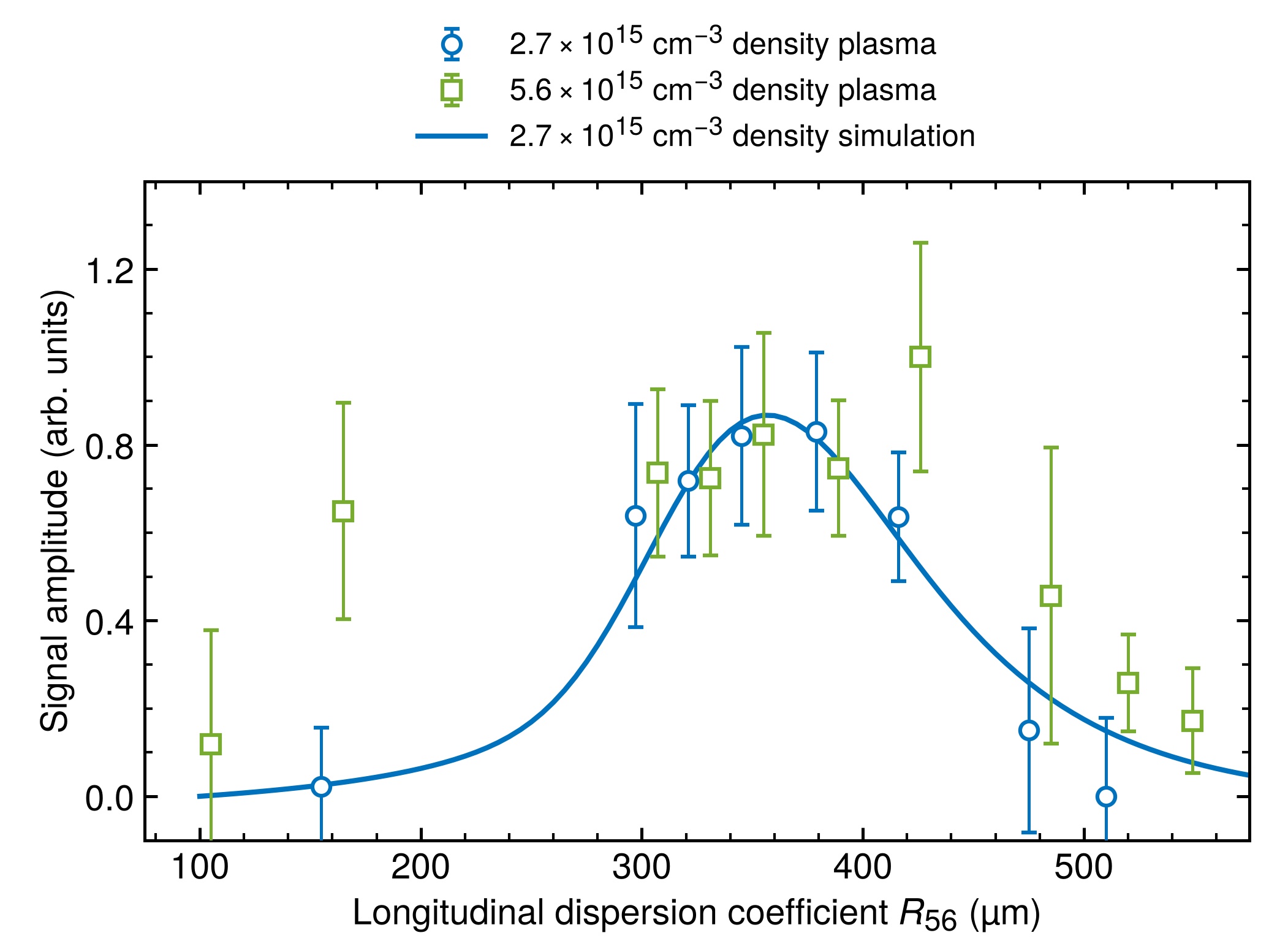}
\caption{\textbf{Transition radiation.} Amplitude of the coherent transition radiation signal measured on a pyroelectric detector downstream of the bunch compression chicane as a function of the chicane longitudinal dispersion coefficient $R_{56}$. Experimental data is shown for plasma densities of $2.7\times10^{15}$~cm$^{-3}$ and $5.6\times10^{15}$~cm$^{-3}$. The error bars correspond to the standard deviation. Solid blue line is a simulation of the CTR signal for an ensemble of linearly chirped electron bunches. Note that $R_{56} = 100\,\mu\mathrm{m}$ corresponds to zero magnetic field in the chicane.}\label{Figure_3}
\end{figure}

The projected energy spread measured on the energy spectrometer provides a measure of the average chirp along the witness bunch. Based on measurements of the average and range of projected energy spreads and our simulations of the witness bunch length at the plasma entrance, we estimate the chirp acquired by the witness bunch at the plasma exit is in the range of $\sim$0.12--0.56~MeV/$\mu$m. 
Such a large energy chirp is used to compress the bunch downstream of the plasma with a small, meter-scale bunch compressor using weakly bending dipoles with peak magnetic field $\sim$20~mT. The ratio between the witness bunch length at the exit and entrance of the bunch compressor (the compression factor) is set by the dipole magnetic fields and the distance between the dipoles and is parametrized by the longitudinal dispersion coefficient $R_{56}$. Maximum compression is achieved when the magnitude of the $R_{56}$ is set to equal the inverse of the bunch chirp at the chicane entrance. 

\par Following the bunch compression chicane, the compressed bunch is incident on an aluminum foil, where it emits coherent transition radiation (CTR) whose total energy scales inversely with the length of the compressed bunch. The CTR signal is measured on a pyroelectric detector (see Methods for details). We performed a scan of the chicane longitudinal dispersion while measuring the total CTR energy to determine the $R_{56}$ resulting in maximum compression, and the results are shown in Fig.~\ref{Figure_3}.  Note that the propagation of the beam through free space from the plasma exit to the chicane entrance adds 100~$\mu$m to the $R_{56}$, such that even when the chicane is fully turned off, the minimal $R_{56}$ is 100~$\mu$m.

In Fig.~\ref{Figure_3},  for both $2.7\times10^{15}$~cm$^{-3}$ and $5.6\times10^{15}$~cm$^{-3}$ densities, we observe that the signal rises initially as $R_{56}$ increases, finds a maximum value, and then decreases. This is consistent with the witness electron bunch compressing, reaching a minimum bunch length and then lengthening as it goes beyond full compression. The simulation curve shows the calculated CTR signal as a function of $R_{56}$ for linearly chirped Gaussian witness bunch of 30~fs RMS duration at the chicane entrance. In order to capture shot-to-shot jitter in our simulation of the multishot measurement data, the calculated CTR signal at each $R_{56}$ value is the average of an ensemble of linearly chirped bunches with a Gaussian distribution centered on 0.29~MeV/$\mu$m with an RMS width of 0.04~MeV/$\mu$m fitted to the $2.7\times10^{15}$~cm$^{-3}$ data. The $2.7\times10^{15}$~cm$^{-3}$ density data agrees well with the simulated result. 

\par In this experiment, control over the plasma density was utilized to exert control over the chirp of the electron beam. A more strongly chirped electron beam will experience a higher correlated energy spread, introducing higher order compression effects and widening the range of chicane $R_{56}$ values where peak compression is achieved \cite{Swanson2026}. This behavior is observed in the $5.6\times10^{15}$~cm$^{-3}$ density case shown in Fig.~\ref{Figure_3}, which exhibits a peak signal amplitude over a broader range of $R_{56}$ values. Based on our estimates of the energy chirp and slice energy spread of the witness beam and our beam dynamics simulations matching the data in Fig.~\ref{Figure_3}, we expect the compression factor to be one order of magnitude and the final witness bunch duration to be fs-scale at the exit of the bunch compressor.

\begin{figure}[t]
\centering
\includegraphics[width=1.0\linewidth]{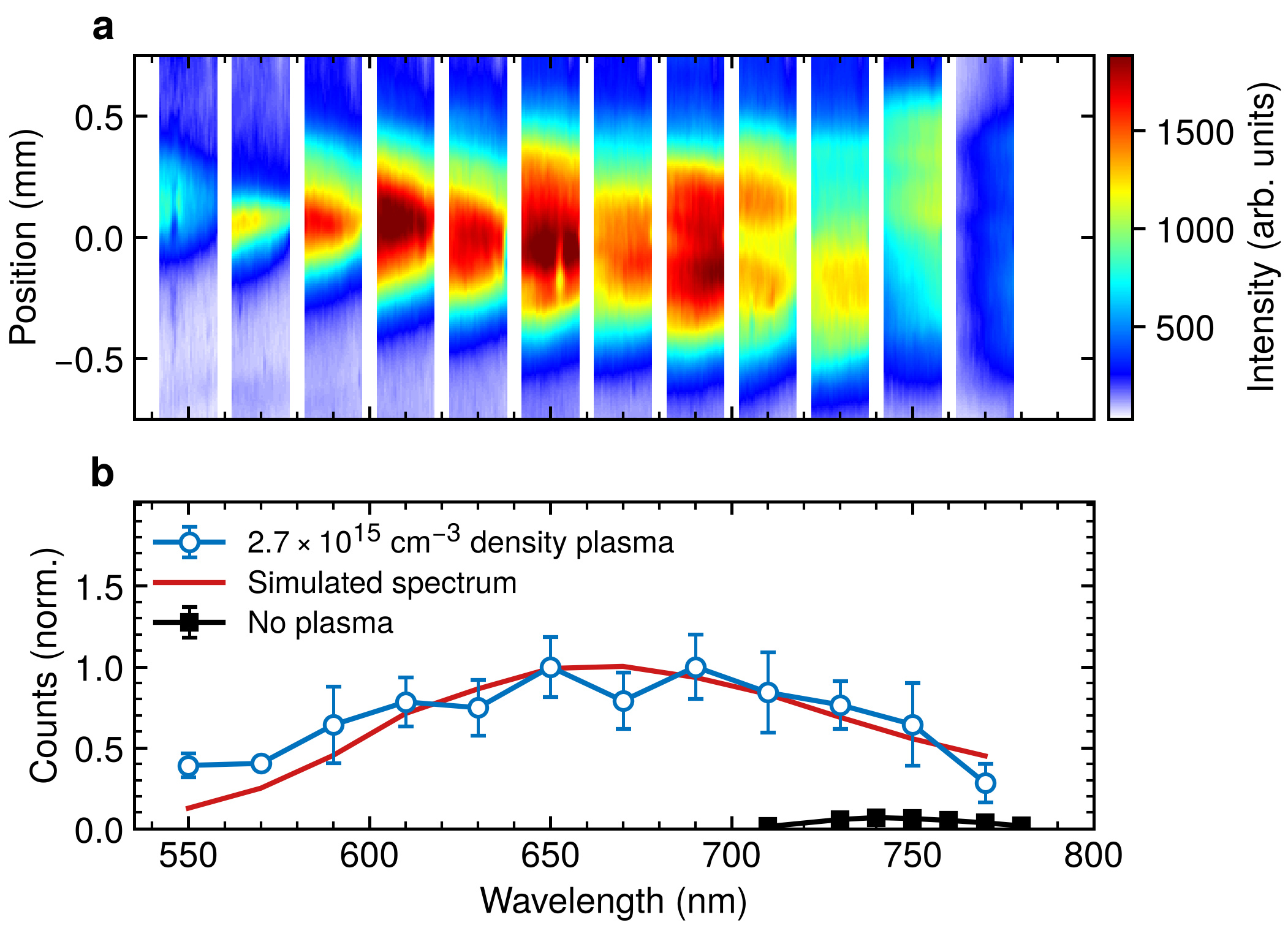}
\caption{\textbf{Undulator radiation.} (a) Coherent undulator radiation spectra measured with a 20~pC beam and $2.7\times10^{15}$~cm$^{-3}$ density plasma at the exit of a single meter-scale undulator. (b) Normalized spectral counts comparing undulator radiation produced with a plasma-compressed beam versus without plasma compression, along with the simulated spectrum (see Methods). Error bars, standard deviation over shots. With plasma, we observe a wide spectrum centered on 665~nm with an RMS bandwidth of 59~nm. This large bandwidth is consistent with a single microbunch coherent emission over a few-period undulator. The black line depicts the undulator radiation spectrum taken with no plasma.}\label{Fig4_undulatorSpectra}
\end{figure}

The fully compressed $\mathcal{O}(\mathrm{fs})$ duration electron beam is then sent into an undulator whose resonant wavelength is tuned to 650~nm and is comparable to the electron bunch length. The electron beam emits coherent undulator radiation as a single microbunch, leading to the generation of a $7.6 \pm 1.3$~nJ photon pulse at the exit of a single undulator module of 77 periods in length, with the pulse energy measured by a photodiode and the spectrum by a grating spectrometer. For comparison, this pulse energy result obtained from a single undulator module is about four orders of magnitude greater than the first demonstration of a beam-driven, plasma-based, self-amplified spontaneous emission (SASE) FEL at SPARC using seven modules and is a result of the higher peak current and pre-bunching of the beam on the scale of the radiation wavelength~\cite{Pompili2022}.

\begin{figure}[!t]
\centering
\includegraphics[width=\linewidth]{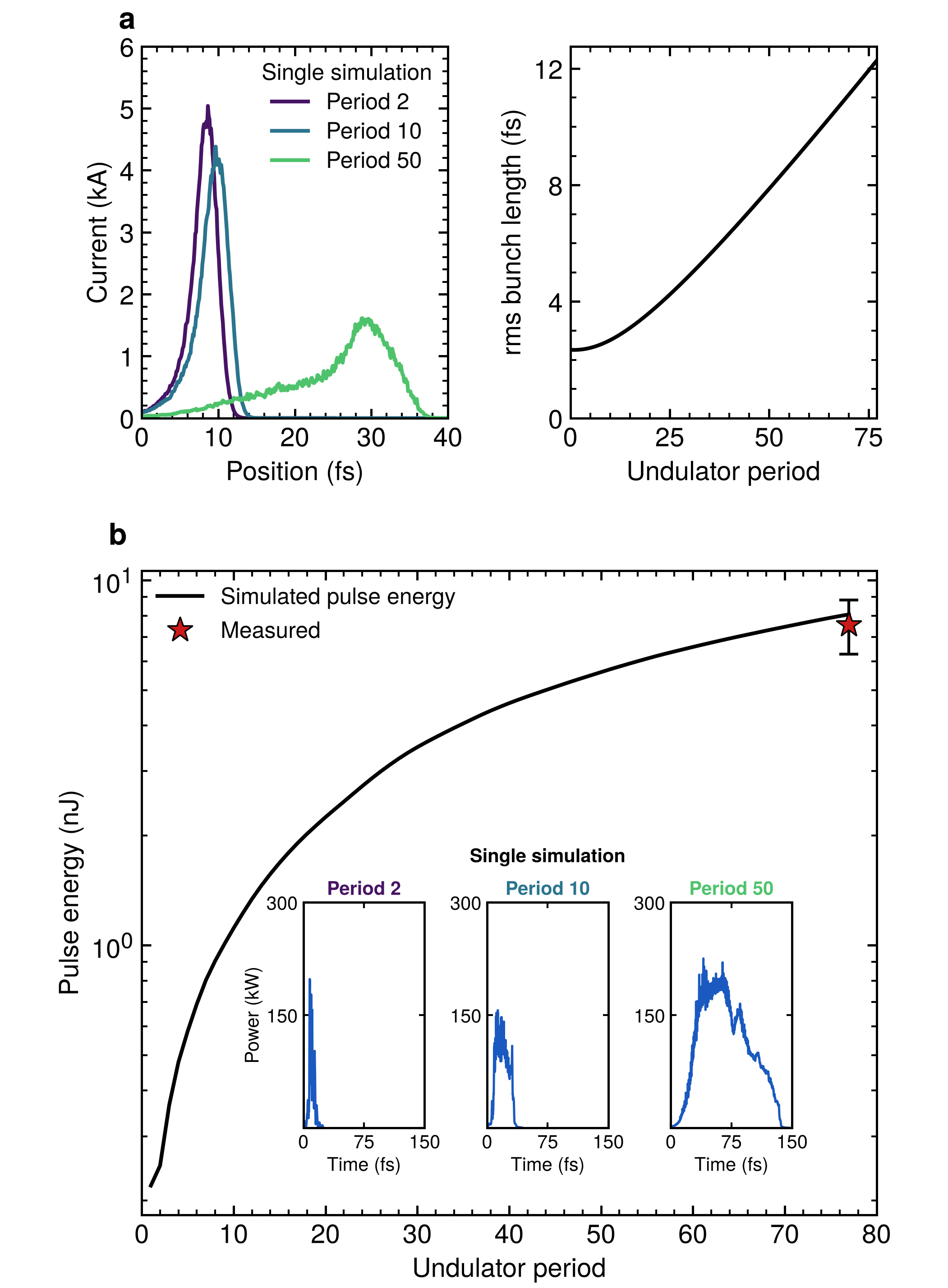}
\caption{\textbf{FEL simulations.} (a) Evolution of the current profile through the undulator, as well as the bunch length, for the 106~MeV, 0.3~MeV/$\mu$m beam, shown at undulator periods 2, 10 and 50. (b) Simulated pulse energy along the undulator, averaged over multiple simulations (see Methods for details). The pulse energy measured with the photodiode, $7.6 \pm 1.3$~nJ (mean $\pm$ standard deviation over shots), is shown on the plot, with insets showing the power profiles from simulation along the undulator.}\label{Figure_5}
\end{figure}

Fig.~\ref{Fig4_undulatorSpectra}b shows a comparison of the radiated spectrum with plasma on and without plasma. Without plasma, the witness bunch does not acquire a significant energy chirp and is therefore not compressed by the chicane. The plasma-off data was taken with the maximum $R_{56}$, 560~$\mu$m, to allow for the most possible compression with the small residual chirp present on the witness bunch following the velocity bunching and acceleration in the linac sections. The measured undulator radiation spectrum with plasma-off stays below $\sim$7\% of the peak signal with plasma. Note that the plasma-off spectrum is centered around 750~nm because the lower energy witness bunch red-shifts the FEL resonance condition relative to the plasma-on case.

\par To interpret the measured undulator radiation spectra, we simulated the single microbunch free electron lasing interaction with the simulation code PUFFIN \cite{Campbell2012}. The simulation results are shown in Fig.~\ref{Figure_5} and described further in the Methods. Simulations were performed using electron beams centered at mean energies of 106~MeV and 109~MeV to match the range of measured energy spectra of the beam produced by the $2.7\times10^{15}$~cm$^{-3}$ density plasma. At each mean electron energy, the simulations consider beams entering the undulator with a range of chirps varying from 0.2 to 0.4~MeV/$\mu$m, which were chosen to match the expected chirps from pyroelectric detector measurements shown in Fig.~\ref{Figure_3}.  Longitudinal space charge effects and the undulator's longitudinal dispersion cause the electron beam phase space to evolve and the bunch to decompress as it transits the undulator. The peak current of the electron bunch falls from 4.8~kA at the undulator entrance to 1.0~kA at the exit, while the RMS bunch length grows from 2.3~fs to 12~fs (Fig.~\ref{Figure_5}a).

The averaged undulator radiation spectra simulated at the two different beam energies fit to the measured data as shown in Fig.~\ref{Fig4_undulatorSpectra}b. Details on the fitting procedure and results are provided in the Methods and in Extended Data Fig.~\ref{fig:current_evolution}. The simulated FEL pulse energy curve in Fig.~\ref{Figure_5}b represents the total energy in the spectrum produced by taking the weighted sum of the spectra. The simulated total pulse energy reaches 8.1~nJ at the undulator exit, within the uncertainty of the $7.6 \pm 1.3$~nJ measured with a photodiode located at the undulator exit. The simulated RMS bandwidth of 8.2\% is also in good agreement with the measured bandwidth of $8.9 \pm 0.3$\%.

\par We have demonstrated, for the first time, compact and extreme compression of a relativistic electron bunch driven by a plasma wakefield. Analogously to chirped pulse amplification for lasers, the wakefield of a preceding drive bunch imprints a large, tunable energy chirp ($\sim$0.12--0.56~MeV/$\mu$m) on a witness bunch within a 13-cm-long plasma discharge. Because the energy chirp is so large in magnitude, a weakly bending, compact chicane ($R_{56}$ up to $\sim$400~$\mu$m) suffices to compress the bunch temporally by one order of magnitude. This compact, extreme compression with weakly bending fields is poised to relax the long-standing trade-off between bunch length and transverse beam quality that stands as a fundamental limitation towards pushing electron bunch compression beyond the current state-of-the-art.

The resulting order femtosecond-duration witness bunch was used to drive single-microbunch free electron lasing at 650~nm in a single meter-scale undulator, yielding a pulse energy of $7.6 \pm 1.3$~nJ, about four orders of magnitude above the first beam-driven, plasma-based SASE FEL demonstration \cite{Pompili2022}. This proof-of-principle experiment establishes plasma-based compression as a compact and scalable route towards few-cycle, waveform-controlled FELs extending to the EUV and soft X-ray regimes~\cite{Emma2021,Hessami2024}, with direct applications in ultrafast quantum and materials science, strong-field quantum electrodynamics and beyond. Future experiments will explore scaling these experimental results to shorter bunch durations and shorter FEL wavelengths via plasma-based injection \cite{Zhang2025, Deng2019} as well as developing multi-pulse multi-color schemes to enable photon based ultrafast pump-probe dynamics studies.

\newpage
\section*{Methods}\label{sec1}

\subsection{Generation of the two electron bunches}\label{generation}
The driver and witness bunches are directly generated on the photocathode by two ultra-short ($\approx 100$~fs) blue laser pulses ($\lambda_L\approx 390$~nm) delayed by 3~ps~\cite{Ferrario2011,pompili2021time}. 
The photocathode laser system is based on a commercial Ti:Sa chirped pulse amplified laser producing $\approx 60$~mJ energy, centered at $\lambda_L\approx 780$~nm central wavelength. The frequency is then doubled with a $\beta$-BBO nonlinear crystal and shaped in energy, transverse and longitudinal sizes.
A split and delay system has been realized to generate multiple laser pulses. It is composed by custom-made polarizing beam splitters (PBS) that reflect vertical polarization and transmit horizontal one. Motorized zero-order half-wave plates rotate the polarization of the incoming pulse in order to control the intensity of each replica. Translation stages in each line allow to control the delay between the pulses. Two motorized irises, positioned at the same distance from the first PBS, are used to control the transverse sizes of the two pulses. 
The pulse train is then transported up to the copper photocathode with several lenses in order to obtain a top-hat transverse profile in correspondence of the cathode. 
The temporal separation and durations of the two bunches are tuned with the velocity-bunching technique that requires injection in the first traveling-wave section at the zero crossing of the RF wave. The longitudinal compression is then achieved by slightly accelerating the tail of the beam while decelerating its head. This leads to a rotation of the beam longitudinal phase-space that is simultaneously chirped and compressed. To avoid uncontrolled emittance growth, solenoids embedded on the linear accelerator are turned on and provide the necessary additional focusing~\cite{ferrario2010experimental}. Unlike other methods, the velocity-bunching does not employ any mask or scraper in a dispersive section, and therefore there is no loss of charge. 

\subsection{Plasma source}\label{plasma}
For the experiment, a 13-cm-long capillary with a 2~mm diameter hole has been used. The capillary is made in LEXAN polycarbonate and has one inlet for the gas injection.
The plasma is produced by ionizing nitrogen gas.
A high-speed solenoid valve, located outside of the vacuum chamber, is used to fill the capillary with the gas. The valve is opened for 2~ms and the discharge current is applied immediately after its closure. The discharge current is generated by a high-voltage generator that provides 9~kV pulses and a $\approx 400$~A current through the capillary. The current is monitored with a Pearson current monitor.
The plasma density is measured with a Stark broadening-based diagnostics measuring the H$_{\beta}$ Balmer lines~\cite{biagioni2019temperature}, and measuring the energy loss of trains of low-charge electron bunches resonantly driving linear plasma wakefields~\cite{Romeo2021,VERRA:2025}.
The stabilization of the discharge process and plasma formation is provided by pre-ionizing the gas with a Quantel 532~nm CFR Ultra Nd:YAG laser, reducing the discharge timing-jitter to the order of a few nanoseconds~\cite{articolo_angelo}.
The laser is installed close to the capillary vacuum chamber and is injected into it by means of a metallic mirror directing the laser at the capillary entrance. The best results in terms of the discharge-timing jitter ($\approx$1--2~ns) are obtained by injecting the laser into the capillary approximately 100~ns before the start of the discharge. Its transverse spot size was set to $\approx 5$~mm with $\approx 200~\mu$J energy.
We measure the time of arrival of the electron bunches with respect to the discharge pulse by using the analog signals coming from a beam position monitor located upstream of the capillary and the Pearson current monitor. The two signals are displayed on a digital scope with 500~MHz bandwidth.

\subsection{Beam focusing and matching in the plasma}\label{MatchingPlasma}
Optimal acceleration without detrimental effects on the emittance requires the witness beam to be transversely matched to the plasma.
The beam focusing at the plasma entrance is obtained by using three PMQs with $\approx 460$~T/m gradient and lengths $L_1=10$~mm and $L_2=L_3=20$~mm. The resulting focal length is $f\approx 5$~cm. The overall focusing of the triplet can be adjusted to match the beam energy by longitudinally moving each PMQ with a nanometer-resolution piezoelectric motor~\cite{pompili2018compact}. A similar triplet of PMQs is also installed downstream of the capillary to allow the capture of the high-divergence accelerated witness.

\subsection{Magnetic chicane layout}\label{Chicane}
The transfer line from the SPARC linac exit to the undulator entrance is 5.75~m long and includes two triplets of electromagnetic quadrupoles, used for matching the beam parameters at the exit of the linac to those of the undulator beamline.
The layout of the transfer line has been realized in order to perform a small orbit in a magnetic chicane. The maximum horizontal deviation of the electron beam orbit with the chicane magnets turned to maximum field strength can reach 12~mm at the center of the chicane. The chicane consists of four bending magnets, which allow for having beam displacement and no angle at the desired injection location. The chicane allows to realize bumps for electron energies up to 200~MeV and uses the spectrometer magnet (operating with small currents) as the first bump corrector.
The following three bending magnets have magnetic lengths of approximately 13~cm.

\subsection{Bunch length measurement using transition radiation diagnostics}\label{sec:methods_pyro}


The spectral intensity of coherent transition radiation from the compressed witness electron beam scales as:

\begin{equation} \label{eqn:coherent_scaling}
    I_{\mathrm{coherent}} \propto N^2 b^2 I_{\mathrm{incoherent}}
\end{equation}

where $b(k) = \frac{1}{N} \sum_n e^{i \vec{k}\cdot \vec{r}_n}$ is the electron bunching factor and $I_{\mathrm{incoherent}}$ is the spectral intensity of transition radiation from a single electron. The pyroelectric sensor (Gentec-EO THZ2I-BL-BNC, 0.25--3000~$\mu$m) operates by measuring a signal proportional to the total energy absorbed, effectively giving the integral of equation~(\ref{eqn:coherent_scaling}) over its bandwidth. Since the total charge in the witness beam is approximately constant shot-to-shot, any variation observed can be attributed to a change in a bunching factor. For a Gaussian bunch with longitudinal duration $\sigma_t$, the bunching factor can be written as:

\begin{equation} \label{bfac}
    b\left(k\right) = e^{-k^2 c^2\sigma_t^2/2}
\end{equation}

illustrating the pyro signal's proportionality to the bunch length of the witness beam. 

In performing the CTR measurements, 50 shots were taken at each $R_{56}$ setting with both the witness and driver bunches, and 50 additional shots were taken with only the driver bunch. For each setting, the outlying shots more than 2.5 robust standard deviations away from the median were removed, and the mean driver-only signal was subtracted from the mean witness-and-driver signal. This difference, shifted and scaled to vary from 0 to 1, represents the CTR signal amplitude due to the witness beam shown in Fig.~\ref{Figure_3}. The error bars give the shot-to-shot standard deviation of the witness signal, found by subtracting the variance of the driver-only shots from the variance of the witness-and-driver shots.

\subsection{Undulator radiation detection and analysis} \label{UndAnalysis}

In the experiment the compressed electron bunch is transported through an undulator magnet and generates coherent undulator radiation. The undulator is a single planar module of 77 periods, with a resonant wavelength given by
\begin{equation}
\lambda_r = \frac{\lambda_u}{2\gamma^2}\left(1+\frac{K^2}{2}\right),
\label{eqn:resonance}
\end{equation}
where $\lambda_u = 28.18$~mm is the undulator period, $\gamma$ is the electron Lorentz factor and $K$ is the undulator parameter, given by
\begin{equation}
K = \frac{e B_0 \lambda_u}{2\pi m_e c},
\label{eqn:Kparam}
\end{equation}
with $B_0$ the peak on-axis magnetic field, $e$ the elementary charge, $m_e$ the electron mass and $c$ the speed of light. For $K = 1.4$ and a beam energy of 106~MeV, the resonant wavelength is approximately 650~nm. The $K$-value was controlled by modifying the gap, where a previously measured map between the gap distance and measured $K$ value was employed.

The coherent radiation produced in the undulator was reflected out of the beamline with a thin aluminum target and sent into a single shot grating spectrometer with a spectral bandwidth of 16~nm and tunable central wavelength. The full spectrum was obtained as a multishot measurement, averaging the individual spectra from multiple shots over each 16~nm range while scanning the spectrometer central wavelength. These spectra are depicted in Fig.~\ref{Fig4_undulatorSpectra}. 

Once individual shots were taken, they were processed by first subtracting the averaged drive-only signal from the drive+witness shots. Then, the shots that were more than 3 robust standard deviations above the median were taken for each setting and averaged. This produced the multishot spectra.

\subsection{Electron beam compression and FEL simulations}\label{FELsim}

The FEL interaction was simulated using the three-dimensional FEL code  PUFFIN ~\cite{Campbell2012}. This FEL code is well suited for simulating FELs that use ultrashort beams due to the fact that the code does not employ the slowly varying envelope approximation nor the period averaging approximation. In addition, to simulate high current beams of duration comparable to the FEL resonant wavelength such as the one considered in this work, the simulations use a custom module that simulates the effects of longitudinal space charge (LSC) which introduce non-negligible dynamics and affect of the beam distribution during its traversal in the undulator
primarily through the generation of energy chirps in the longitudinal phase space \cite{Hessami2024}. 

The FEL simulations consider 20~pC charge Gaussian beams at energies of 106 and 109~MeV with an RMS bunch duration of 30~fs at the exit of the plasma and a 0.1\% uncorrelated energy spread with linear energy chirps ranging between 0.20 and 0.40~MeV/$\mu$m. The beams were propagated through free space with an $R_{56}$ value of 100~$\mu$m, compressing them through a chicane with an $R_{56}$ value of 250~$\mu$m, and then transporting those variously compressed beams into a $K = 1.4$, 77-period undulator simulated with PUFFIN. This approach was employed to mimic the shot-to-shot variation in chirp and mean energy of the electron bunches encountered in the experiment. At each chirp, the simulated spectra of the two beam energies, integrated over the full transverse simulation grid, were averaged to represent a $106 \pm 3$~MeV beam. The weights of the five chirps were then found by a non-negative least-squares fit to the measured spectrum. In Extended Data Fig.~\ref{fig:current_evolution}, the different colors in the fill represent the contributions of the different chirps, with the red line representing the sum total simulated spectrum. Fig.~\ref{Figure_5}b shows the pulse energy along the undulator averaged with the same weights, which reaches 8.1~nJ at the undulator exit, compared with the $7.6 \pm 1.3$~nJ experimentally measured with a photodiode located at the undulator exit.

The pyro signal was simulated by generating a Gaussian witness beam with a bunch length of 10~$\mu$m, 20~pC, an energy of 106~MeV, and an uncorrelated energy spread of 0.1\% with a linear chirp varying from 0.2 to 0.4~MeV/$\mu$m. The beam was then compressed through the chicane, modeled as a longitudinal map $z_f = z_i + R_{56}\delta_E + T_{566}\delta_E^2$, and the coherent transition radiation emitted by the compressed beam at the aluminum foil was calculated. This was performed for a range of chirps and the signal was averaged over a Gaussian distribution of chirps where the mean and width were fit to the data, giving a mean chirp of 0.29~MeV/$\mu$m and an RMS width of 0.04~MeV/$\mu$m.

\subsection{Radiation diagnostics}\label{FELdiagn}

\par The energy of the radiation is measured by means of a silicon photodiode installed downstream of the undulator. The radiation is extracted by inserting an in-vacuum high-reflective mirror. The photodiode (Thorlabs PDA100A), whose response was previously calibrated using a reference 800~nm laser, giving a calibration of 2.46~nJ/V, collects the light through a focusing lens, and their signals are read with a digital scope (LeCroy WS454).

\par Here, the FEL radiation is also transported to the imaging spectrometer (Acton Spectra Pro).
An imaging system consisting of several high-reflective mirrors and a lens with 15~cm focal length make a magnification $1:1$ of the FEL radiation transverse size at the spectrometer entrance slit.
The spectrometer has a diffraction grating with 1200 lines/mm and can be rotated to accommodate wavelengths in the 200--900~nm spectral range.

\par Adopting a CCD coupled with a Hamamatsu C9547-01 intensifier (ICCD), we image the dispersed light~\cite{Pompili2022,GALLETTI:2022a,Galletti2026}. The ICCD camera is positioned at the exit of the imaging spectrometer. 

\par The ICCD gate duration is set to 5\,ns with an available scanning delay with respect to the pulse arrival time. The dispersed light is measured with a spectral resolution of 0.06~nm per pixel. The entire spectral window available with such a configuration is about 16~nm.

\subsection*{Data availability}
The data that support the findings of this study are available from the corresponding authors upon reasonable request.

\subsection*{Code availability}
The analysis and simulation codes used in this study are available from the corresponding authors upon reasonable request.

\noindent

\backmatter

\bmhead{Acknowledgments}
We thank I. Balossino, M. Ferrario, S. Romeo, A. Zigler, A. Michelotti, A. Vannozzi, A. Liedl, L. Giannessi, and all the machine operators involved in the experimental run.

\section*{Authors Contributions}
C. E., K.S, R. H., M.G., R.P., and L.V. planned and managed the experiment with inputs from all the co-authors. M.G., and L.V., managed the photocathode laser system. L.C., and A.B. designed and characterized the capillary system. C. E., K.S, R. H., M.G., R.P., and L.V. carried out the experimental campaign, performed the data analysis and numerical simulations. A.M. and C.E. conceptualized the experiment. The manuscript was written by C. E., K.S, R. H., M.G., R.P., and L.V. with inputs from all the co-authors.  All authors extensively discussed the results and reviewed the manuscript.

\section*{Competing interests}
The authors declare no competing interests.

\section*{Declarations}

This work has received funding from the United States Department of Energy Early Career Research Program Award.

\bibliography{MainBibliography}

\clearpage
\section*{Extended Data}
\setcounter{figure}{0}
\renewcommand{\figurename}{Extended Data Fig.}

\begin{figure}[h]
\centering
\includegraphics[width=\linewidth]{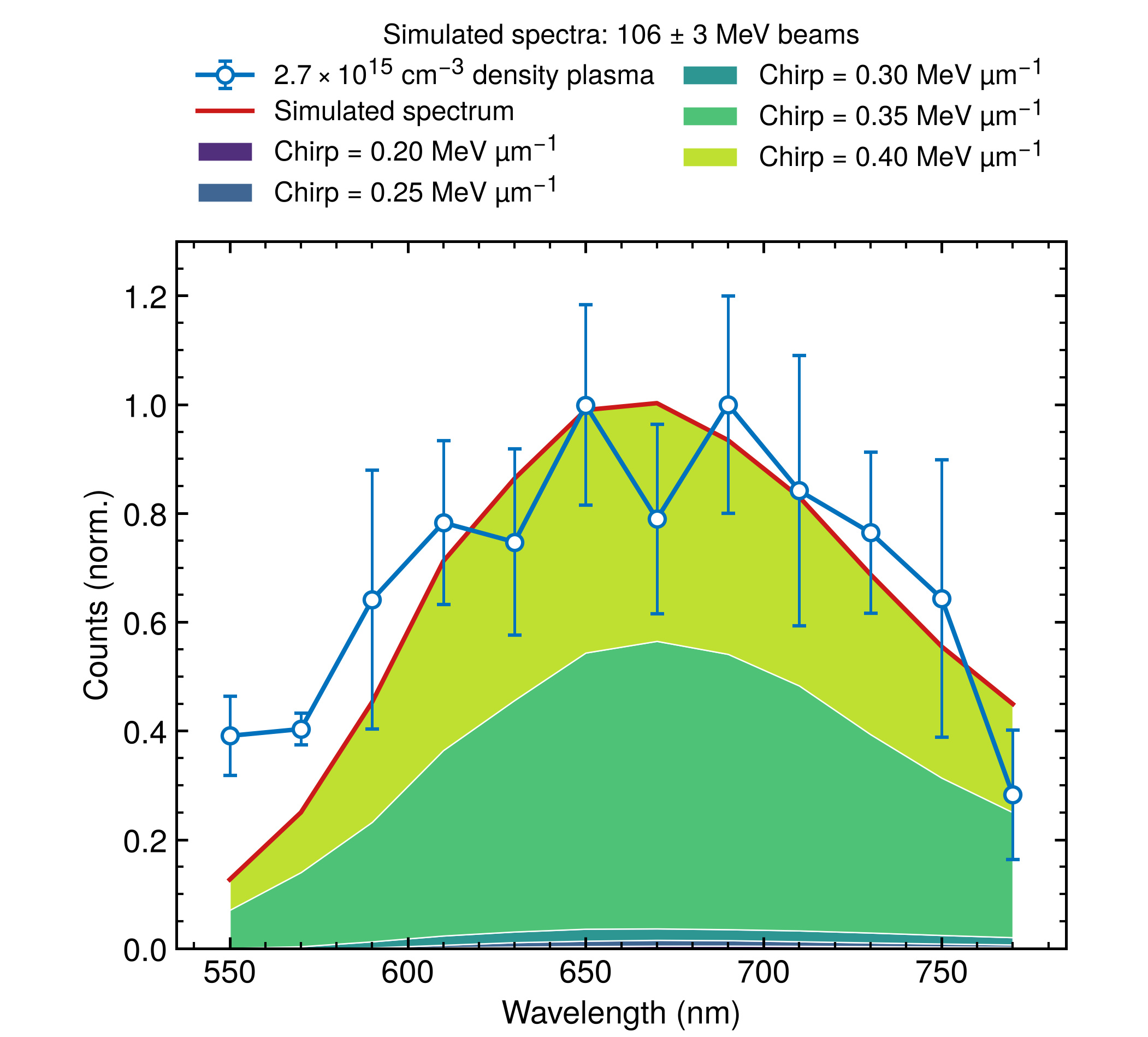}
\caption{\textbf{Undulator radiation simulations.} Measured multishot undulator spectrum compared with a weighted sum of simulated spectra for 20~pC charge, 30~fs RMS duration beams of $106 \pm 3$~MeV with linear chirps of 0.20--0.40~MeV/$\mu$m, compressed at the same $R_{56}$ as the measured spectrum. Each colored band is the contribution of one chirp. Error bars, s.d. over shots.}\label{fig:current_evolution}
\end{figure}

\end{document}